\documentclass[12pt]{article}
\usepackage{graphicx}
\usepackage{epsfig}
\usepackage{epstopdf}
\DeclareGraphicsExtensions{.pdf,.eps,.png,.jpg,.mps}

\def\lsim{\mathrel{\rlap{\lower3pt\hbox{\hskip0pt$\sim$}}
     \raise1pt\hbox{$<$}}}         
\def\gsim{\mathrel{\rlap{\lower4pt\hbox{\hskip1pt$\sim$}}
     \raise1pt\hbox{$>$}}}         

\usepackage{amsmath}
\usepackage{amsfonts}

\begin{document}
\begin{titlepage}

\centerline{\Large \bf Quantization Rules for Dynamical Systems}
\medskip

\centerline{Zura Kakushadze$^\S$$^\dag$\footnote{\, Zura Kakushadze, Ph.D., is the President of Quantigic$^\circledR$ Solutions LLC,
and a Full Professor at Free University of Tbilisi. Email: \tt zura@quantigic.com}}
\bigskip

\centerline{\em $^\S$ Quantigic$^\circledR$ Solutions LLC}
\centerline{\em 1127 High Ridge Road \#135, Stamford, CT 06905\,\,\footnote{\, DISCLAIMER: This address is used by the corresponding author for no
purpose other than to indicate his professional affiliation as is customary in
publications. In particular, the contents of this paper
are not intended as an investment, legal, tax or any other such advice,
and in no way represent views of Quantigic$^\circledR$ Solutions LLC,
the website \underline{www.quantigic.com} or any of their other affiliates.
}}
\centerline{\em $^\dag$ Free University of Tbilisi, Business School \& School of Physics}
\centerline{\em 240, David Agmashenebeli Alley, Tbilisi, 0159, Georgia}
\medskip
\centerline{(March 6, 1992; revised: May 10, 2015)\footnote{\, This note, with some (primarily, linguistic) differences, appeared on March 6, 1992 as a preprint (CLNS 92/1137) of Newman Laboratory of Nuclear Studies at Cornell University, where I was a graduate student at the time. A scanned version of the preprint is available from the KEK library: http://ccdb5fs.kek.jp/cgi-bin/img/allpdf?199203193
}}

\bigskip
\medskip

\begin{abstract}
{}We discuss a manifestly covariant way of arriving at the quantization rules based on causality, with no reference to Poisson or Peierls brackets of any kind.
\end{abstract}
\medskip

\end{titlepage}

\newpage
\section{Introduction}

{}The canonical quantization of dynamical systems replaces classical dynamical variables by operators, and classical Poisson brackets by commutators, such that the Correspondence Principle is satisfied. This procedure comes with some ``drawbacks''. On the one hand, it is not manifestly covariant, which is unappealing in relativistic theories. On the other hand, it masks the importance of measurement in quantum theory. These ``shortcomings'' can be circumvented via Peierls brackets \cite{Peierls}, a manifestly covariant generalization of Poisson brackets. In this approach the roles of elementary and complete measurements in quantum theory are prominent \cite{DeWitt}.

{}In this note we discuss a way of arriving at the quantization rules based on the Causality Principle, with no reference to Poisson or Peierls brackets of any kind. We use DeWitt's condensed notations \cite{DeWitt}. We focus on bosonic theories. A generalization to superclassical systems with Grassmann valued variables is straightforward.

\section{Classical Fields}

{}Let $S[\phi]$ be a real local action functional for a classical dynamical system described by a set of real variables $\phi^i$. The classical dynamical equations of motion read:
\begin{equation}\label{cl.eq}
 S_{,i}[\phi] = 0
\end{equation}
Here $i$ is a generic index, which combines a discrete label for the field components and a continuous label for the spacetime points which the field $\phi^i$ depends on. The left hand side of (\ref{cl.eq}) is the first functional derivative of $S[\phi]$. Repeated indices imply summation and integration. We will omit the arguments of classical functionals; thus, $S$ stands for $S[\phi]$, where $\phi^i$ is an arbitrary solution of (\ref{cl.eq}) so long as it is not a singular point of the functional $S$.

{}Consider the case where the continuous matrix $S_{,ij}$ is nonsingular, i.e., there is no constraint and, therefore, the action does not possess any infinite-dimensional invariance group. Here the following remarks are in order. First, the notion of a constraint is understood in the context of a ``gauge algebra'' \cite{Batalin}. Second, while for a finite matrix nonsingularity implies that it has no null eigenvalue, for continuous matrices the notion of eigenvectors and eigenvalues is subtle. Thus, the equation
\begin{equation}\label{null}
 S_{,ij}~f^i = 0
\end{equation}
has nontrivial solutions even if $S_{,ij}$ is nonsingular. As a general rule, a continuous matrix can be considered nonsingular if it has no eigenvector with a null eigenvalue, either vanishing outside a limited region of spacetime, or quadratically integrable. In (\ref{null}) $f^i$ satisfies neither of these conditions.

{}Since $S_{,ij}$ is a nonsingular matrix, it can be inverted. The inverse depends on boundary conditions. For example, the advanced and retarded Green functions $G^{+ij}$ and $G^{-ij}$ satisfy the following equations and boundary conditions:
\begin{eqnarray}
 &&S_{,ik}~ G^{+kj} = -{\delta_i}^j;~~~G^{+ij} = 0,~~~i > j\label{ad}\\
 &&S_{,ik}~ G^{-kj} = -{\delta_i}^j;~~~G^{-ij} = 0,~~~j > i\label{ret}\\
 &&G^{+ij} = G^{-ji}\label{adret}
\end{eqnarray}
Here the delta-symbol ${\delta_i}^j$ is understood to contain a spacetime delta-function. The symbol ``$>$'' means ``is in the future with respect to".

\section{Operators}

{}Quantization amounts to replacing the classical real-valued variables $\phi^i$ by Hermitian operators $\Phi^i$, which, in general, do not commute. Therefore, ambiguities arise in the quantum dynamical equations of motion
\begin{equation}
 S_{,i}[\Phi] = 0
\end{equation}
which need not have the classical form. These ambiguities in products of operators must be resolved by means of their symmetrization, i.e., a real functional $Z^{ij}$ must exist such that
\begin{equation}
 \left[\Phi^i, \Phi^j\right] = i~Z^{ij}[\Phi]
\end{equation}

{}Consider a linear theory described by the action
\begin{equation}
 \Sigma = {1\over 2}~S_{,ij}~\Phi^i~\Phi^j
\end{equation}
and the commutation relations
\begin{equation}\label{Omega}
 \left[\Phi^i, \Phi^j\right] = i~\Omega^{ij}[\Phi]
\end{equation}
In this theory there is no ambiguity in the quantum dynamical equations of motion as they are linear:
\begin{equation}\label{lin}
 S_{,ij}~\Phi^j = 0
\end{equation}
{}From (\ref{Omega}) and (\ref{lin}) we get
\begin{equation}\label{SOmega}
 S_{,ik}~\Omega^{kj}[\Phi] = 0
\end{equation}
It then follows that the functional $\Omega^{ij}$ does not depend on $\Phi^i$ or else Eq. (\ref{SOmega}) would be a constraint, which would contradict our prior assumptions. So, we have:
\begin{equation}\label{SOmega1}
 S_{,ik}~\Omega^{kj} = 0,~~~\Omega^{ij} = -\Omega^{ji}
\end{equation}
where $\Omega^{ij}$ must be constructed solely from $S_{,ij}$ and/or its inverse operators, and we conclude that it is a linear combination of the real Green functions of $S_{,ij}$.

\section{Action Variations}

{}Consider an infinitesimal variation in the functional form of the action:
\begin{equation}\label{deltaS}
 S \rightarrow S + \delta S
\end{equation}
where $\delta S$ vanishes outside a limited region of spacetime. Such a variation can be thought of as describing a measurement process in ``quantum system + macro apparatus'' (see \cite{DeWitt} for details). Then the new dynamical equations of motion
\begin{equation}
 S_{,ij}~\delta\Phi^j = -\delta S_{,ij}~\Phi^j
\end{equation}
must be solved assuming {\em retarded} boundary conditions in accordance with the Causality Principle, i.e.,
\begin{equation}
 \delta\Phi^i = G^{-ij}~\delta S_{,jk}~\Phi^k
\end{equation}
Therefore,
\begin{equation}\label{deltaOmega}
 \delta\Omega^{ij} = -i\left(\left[\delta\Phi^i,\Phi^j\right] + \left[\Phi^i,\delta\Phi^j\right]\right) = G^{-ik}~\delta S_{,kl}~\Omega^{lj} + \Omega^{ik}~\delta S_{,kl}~G^{+lj}
\end{equation}
where we have used (\ref{adret}).

{}Eq. (\ref{deltaOmega}) implies that $\Omega^{ij}$ has a definite transformation property under the action variations (\ref{deltaS}). On the one hand, we concluded in the previous section that $\Omega^{ij}$ is a linear combination of the real Green functions. On the other hand, there are only two real inverse matrices, namely, $G^{\pm ij}$, with definite transformation properties determined by their kinematics (\ref{ad}) and (\ref{ret}):
\begin{equation}\label{deltaG}
 \delta G^{\pm ij} =G^{\pm ik}~\delta S_{,kl}~G^{\pm lj}
\end{equation}
Therefore, $\Omega^{ij}$ must be a linear combination of $G^{\pm ij}$. Taking into account (\ref{SOmega1}), (\ref{deltaOmega}) and (\ref{deltaG}), we get
\begin{equation}
 \Omega^{ij} = \alpha \left(G^{+ ij} - G^{- ij}\right)
\end{equation}
where $\alpha$ is a constant and does not depend on the functional form of $S$. So, we have the following commutation relations:
\begin{equation}
 \left[\Phi^i,\Phi^j\right] = i~\alpha \left(G^{+ ij} - G^{- ij}\right)
\end{equation}
To match the experimental data, $\alpha$ must be the Planck's constant $\hbar$.

\section{Concluding Remarks}

{}The above argument, which employs neither Poisson nor Peierls brackets, can be generalized to constrained systems along the lines of \cite{DeWitt}, and also to interacting nonlinear systems along the lines of \cite{DeWitt1}. For a recent discussion on quantization of non-Lagrangian systems, see, e.g., \cite{Sharapov} and references therein.

\end{document}